\documentclass[12pt, a4paper]{article}
\pdfoutput=1

\usepackage{amsmath}
\usepackage{amsfonts}
\usepackage{amssymb}
\usepackage{graphicx, rotating}
\usepackage{epsfig}
\usepackage{latexsym}
\usepackage{graphicx}
\usepackage{color}
\usepackage{amsmath,bm,amssymb}
\usepackage{cite}
\usepackage{slashed}
\usepackage{hyperref}
\hypersetup{colorlinks, citecolor=bluscuro, linkcolor=black, urlcolor=bluscuro}
\definecolor{rossos}{cmyk}{0,1,1,0.55}
\definecolor{bluscuro}{rgb}{0.15, 0.2, .85}
\definecolor{bluchiaro}{cmyk}{1,.3,0.,0.1}

\newcommand{\lp}{\left(}
\newcommand{\rp}{\right)}
\newcommand{\nn}{\nonumber}

\newcommand{\be}{\begin{equation}}
\newcommand{\ee}{\end{equation}}
\newcommand{\bea}{\begin{eqnarray}}
\newcommand{\eea}{\end{eqnarray}}

\def\SP{Schr\"odinger-Poisson}

\begin{document}

\begin{titlepage}
\begin{flushright}
DESY 17-162, TUM-HEP 1102/17\\
\end{flushright}
\vspace{.3in}

\vspace{1cm}
\begin{center}
{\Large\bf\color{black}
Gravitational collapse in the \\[0.5ex] \SP{} system}\\
\bigskip\color{black}
\vspace{1cm}{
{\large M.~Garny$^a$ and T.~Konstandin$^b$}
\vspace{0.3cm}
} \\[7mm]
{\it $^a$ {Technische Universit\"at M\"unchen, James-Franck-Str. 1, 85748 Garching, Germany}}\\
{\it $^b$ {DESY, Notkestr. 85, 22607 Hamburg, Germany}}
\end{center}
\bigskip

\vspace{.4cm}

\begin{abstract}

We perform a quantitative comparison between N-body simulations and the Schr\"o\-dinger-Poisson system in 1+1 dimensions. 
In particular, we study halo formation with different initial conditions. We observe the convergence of various observables in the Planck constant $\hbar$ and also test virialization. We discuss the generation of higher order cumulants of the particle distribution function which demonstrates that the \SP{} equations should not be perceived as a generalization of the dust model with quantum pressure but rather as one way of sampling the phase space of the Vlasov-Poisson system -- just as N-body simulations. Finally, we quantitatively recover the scaling behavior of the halo density profile from N-body simulations.

\end{abstract}
\bigskip

\end{titlepage}

\section{Introduction \label{sec:intro}} 

The observation of baryonic acoustic oscillations (BAOs) in galaxy distributions~\cite{Percival:2001hw,Cole:2005sx,Eisenstein:2005su} was a milestone of modern cosmology. The use of BAOs as a standard ruler is undisputed and will provide invaluable information on the expansion history of the Universe. 
In addition, future galaxy clustering and lensing surveys will also provide detailed information on the shape of the correlation function and the power spectrum
with percent-level accuracy. This can potentially reveal detailed information on the cosmological model and its parameters, and for example be important 
to discriminate among various extensions of the $\Lambda$CDM model. In order to optimally exploit these data sets it is desirable to
improve and scrutinize our theoretical understanding of all sources of uncertainty, starting from non-linear clustering of dark matter (DM), that forms the basis for an
accurate description.
N-body simulations of DM~\cite{Horner:1960,Aarseth:1979,Efstathiou:1985re,Davis:1985rj,Frenk:2012ph} are playing the major role in large scale structure formation and provide impressive results. Still, systematic errors in the predicted power spectrum~\cite{Schneider:2015yka} might have to be improved in view of the accuracy of upcoming surveys. In addition, the N-body technique by definition samples regions of high matter density very precisely, while underdense regions are sampled less accurately. For certain applications these can still be relevant, e.g. for the distribution and shape of voids, see e.g. \cite{Sutter:2014oca, Chan:2014qka, Hamaus:2016wka, Mao:2016faj}, and for extracting higher moments of the DM distribution such as the velocity power spectrum \cite{Hahn:2014lca}.
The observational advance motivates to develop additional independent tools to assess gravitational clustering on BAO scales.

In this work, we consider the \SP{} system as an alternative to N-body simulations~\cite{Widrow:1993qq}. In this framework, the Planck constant $\hbar$ is not a physical quantity. The wave equation describes wave packets that follow the gravitational dynamics and $\hbar$ delineates the smallest unity in phase space in accordance with the uncertainty principle. Ideally, we would like to take the limit $\hbar \to 0$, but reducing $\hbar$ implies more demanding numerical simulations. This picture is very much in the spirit of the original work by Widrow and Kaiser~\cite{Widrow:1993qq} that used the \SP{} system to describe DM for the first time. 

Often, the \SP{} system is used to describe so-called scalar field DM or fuzzy DM~\cite{Turner:1983he,Press:1989id,Sin:1992bg,Goodman:2000tg,Hu:2000ke,Peebles:2000yy,Amendola:2005ad,Chavanis:2011uv,Magana:2012ph,Schive:2014dra,Schive:2014hza,Schive:2015kza,Chen:2016unw,Schwabe:2016rze,Hui:2016ltb}.
In these frameworks, the Planck constant is taken to be physical such that the corresponding de Broglie wavelength is of astronomical size. 
Fuzzy DM may address some of the small-scale issues of the cold dark matter paradigm, such as the core/cusp problem as well as the missing satellites problem \cite{Hu:2000ke, Peebles:2000yy} (see e.g.~\cite{Hui:2016ltb} for a recent discussion). 

The \SP{} system can be perceived as a modification to the dust model via the Madelung representation~\cite{Madelung:1926}. The notable difference is then a quantum pressure term that prohibits the formation of structures below the de Broglie wavelength. At the same time, the evolution of the \SP{} system populates the higher order cumulants after shell crossing~\cite{Uhlemann:2014npa} such that it has the required complexity to be equivalent to the full solutions of the Vlasov-Poisson equations or at least N-body simulations. In the literature, the \SP{} system has often been advocated as an N-body double~\cite{Widrow:1993qq, Uhlemann:2014npa, Briscese:2016ppl} but in fact there are good arguments why the \SP{} system can be directly understood as the coarse grained limit of the Vlasov-Poisson system~\cite{Mauser}. Physically, the classical limit
$\hbar\to 0$ can be understood within the Schwinger-Keldysh formalism of non-equilibrium quantum field theory~\cite{Schwinger:1960qe, Keldysh:1964ud}.
In this formalism, the equation of motion describes the evolution of n-point correlation functions.
After a Wigner transformation~\cite{KB}, these equations reduce in the classical limit  $\hbar\to 0$ (and under the quasi-particle assumption) to the Boltzmann equations. Analogously, two-point functions of the Schr\"odinger wave function in Wigner space are expected to correspond to the particle distribution function for $\hbar\to 0$. The main difference is that the \SP{} system only describes the wave function of a one-particle state while in the Schwinger-Keldysh formalism the scalar field is still an operator acting on the quantum state of the system~\footnote{For a second quantized approach to structure formation see~\cite{Prokopec:2017ldn}.}. We therefore expect that when two wave packets come close, the gravitational softening expressed in the \SP{} system through the quantum pressure term does not actually reflect the true dynamics. Still, in the limit $\hbar \to 0$ also this difference to the full system should cease.

The Madelung representation suggests that the \SP{} system coincides with a pressureless perfect fluid for $\hbar\to 0$, because the quantum pressure term
vanishes. However, this naive expectation is jeopardized by singularities of the Madelung representation of the wave-function, that occur whenever the
density vanishes at a certain time and place. This typically happens once shell-crossings start to play a role. The \SP{} equations themselves are
free of these singularities, and therefore can be integrated also after shell-crossings occur. Therefore, the argument above suggests that the
limit $\hbar\to 0$ needs to be taken with special care within the multi-streaming regime.

The main aim of the present work is to scrutinize the hypothesis that the \SP{} system is actually a double of the Vlasov-Poisson system. To do so, we consider the simplest setting of halo formation in 1+1 dimensions. For this case, the Zel'dovich approximation is exact up to shell crossing, and becomes singular once the first shell-crossing occurs. Detailed N-body simulations have been performed in this case~\cite{Binney:2003sn, Schulz:2012jd} and we will compare our numerical solutions of the \SP{} system with these results.

The structure of the paper is as follows: In Section~\ref{sec:basics} we set up our notation and explain how the \SP{} system is solved numerically. Then, in Sec.~\ref{sec:results} we present our numerical findings before we summarize in Sec.~\ref{sec:disc}.

\section{The \SP{} equations \label{sec:basics}} 

We solve the Schr\"odinger equation
\be
i \hbar \partial_t \psi = - \frac{\hbar^2}{2 a^2 m} \Delta \psi + m V \psi \, ,
\ee
where the gravitational potential $V$ is given by the Poisson equation
\be
\Delta V = \frac{4 \pi G \rho_0}{a} \lp |\psi|^2  - 1  \rp \, . 
\ee
In the case of small $\hbar$, these equations reduce for the Madelung representation
\be
\label{eq:Madelung}
\psi = \sqrt{\rho} \exp \lp i \phi/\hbar\rp \, ,
\ee
and with the identification $\vec u  = \nabla \phi/m$ to the fluid equations
\bea
\label{eq:dust_eom}
\partial_t \rho &=& -\frac{1}{a^2} \nabla (\rho \vec u) \, , \nn \\
\partial_t \vec u &=& - \frac{1}{a^2} \lp \vec u \cdot \nabla \rp \vec u - \nabla V +\frac{\hbar^2}{2 a^2 m^2} \nabla 
\lp \frac{\Delta \sqrt{\rho}}{\sqrt{\rho}} \rp \, .
\eea
Unlike the case of fuzzy DM~\cite{Turner:1983he,Press:1989id,Sin:1992bg,Goodman:2000tg,Hu:2000ke,Peebles:2000yy,
Amendola:2005ad,Chavanis:2011uv,Magana:2012ph,Schive:2014dra,Schive:2014hza,
Schive:2015kza,Chen:2016unw,Schwabe:2016rze,Hui:2016ltb}, we understand the \SP{} equations as a tool to solve the fluid equations (or, ultimately, the Vlasov-Poisson equations) in the limit $\hbar \to 0$. However, there is a crucial difference discussed e.g. in Ref.~\cite{Uhlemann:2014npa}: The fluid equations display an ambiguity when $\rho \to 0$, which happens typically once shell-crossings occur. This ambiguity is absent in the \SP{} system. Hence, it is expected that the \SP{} capture the effects of shell crossing better than the fluid representation. This is also supported by the relation between the Wigner transform of two-point functions and the Boltzmann equation discussed in the introduction and the original work~\cite{Widrow:1993qq, Szapudi:2002cr, Johnston:2009wz}, as well as the general arguments provided in \cite{Mauser}.

The 1+1 dimensional particle distribution function is defined using the Wigner transformation 
\be
f(x,p) = \int  \, \frac{dr}{2 \pi \hbar} \, \psi^*(x + r/2) \, 
\exp \lp \frac{i \, p \, r}{\hbar} \rp  \, \psi(x -r/2) \, ,
\ee
and the first few moments of the particle distribution function are in turn given by the density
\be
\label{eq:defrho}
\rho = \psi^* \psi \, , 
\ee
the fluid velocity field $u$
\be
\label{eq:defu}
u \, \rho \equiv j \equiv \frac{i \hbar}{2} \left[ (\nabla \psi^*) \psi  - \psi^* (\nabla \psi) \right]\, ,  
\ee
and the stress tensor $T$
\be
\label{eq:defsigma}
T \equiv (u^2 + \sigma) \rho \equiv-\frac{\hbar^2}{4} 
\left[(\Delta \psi^*) \psi  - 2 (\nabla \psi^*)(\nabla \psi) + \psi^* (\Delta \psi) \right]\, . 
\ee

In the case of an expanding Universe with scale factor $a(t)$ and Hubble parameter $H = \dot a/a$, one can rewrite the \SP{} system as an evolution equation in  $\eta = \log a(t)$ in the form 
\bea\label{eq:senum}
i \partial_\eta \psi &=& - \frac{\kappa}{2 } \Delta \psi + \bar V \psi \, , \nn \\
\Delta \bar V &=& \frac{3}{2 \kappa} \lp |\psi|^2  - 1  \rp \, , 
\eea
with
\be
\kappa(\eta) = \frac{\hbar}{a^2 m H} \, , \quad
\bar V = \frac{m \, V}{\hbar \, H} \, , \quad 
H^2  = \frac{8\pi G}{3 a^3} \rho_0 \, ,
\ee
where we assumed a matter dominated cosmology with $H^2 \propto a^{-3}$. The function $\kappa(\eta)$ then decreases during the expansion of the Universe.

On the other hand, for a static Universe 
the evolution equation can be formally written in the same form as in (\ref{eq:senum}), when identifying $\eta\to t H$ and $\kappa\to\hbar/(m  H)$ 
where $H \equiv 8\pi G\rho_0/3$ is a constant related to the free-fall time. We stress that this is a formal correspondence, and $H$ is not related
to the Hubble rate in the static case, which instead vanishes.
In the numerical analysis, the only difference between the two cases is that the function $\kappa$ is constant. We also performed simulations in the expanding case to check that our numerical solution reproduces the analytically known Zel'dovich solution up to the first shell-crossing. However, since in comoving coordinates the physical grid spacing increases quickly, static simulations are much more accurate at late times, and we therefore focus on the static case in the following.

There are several approaches in the literature how to solve the Schr\"odinger equation numerically. On the one hand side, one would like to have a discretized equation of motion that is an implicit finite-differencing scheme. On the other hand, unitarity and stability of the solutions are a concern. A common method is Cayley's method that has been used for example in~\cite{Widrow:1993qq,Coles:2002sj}. In the present work, solving the Poisson equation will involve several Fourier transformations such that a run time of order $N \log N$ and not $N$ ($N$ is the size of the discretizing grid) is no concern for us. We use a method that is implicit and stable and very similar to the one used in~\cite{Woo:2008nn}.

Notice that the potential term is local in coordinate space while the kinetic term is local in Fourier space. Hence, whenever only one term is present, the Schr\"odinger equation can be trivially solved by integration resulting in space (or wavemode) dependent exponential factors. To be specific,
\be
\psi(\eta + d\eta,x) = \psi(\eta,x)\exp \left [ -i \, \int_\eta^{\eta+d\eta} d\eta'  \,\bar V(\eta',x) \right]  \, ,
\label{eq:Vapprox}
\ee
and 
\be
\psi(\eta + d\eta,k) = \psi(\eta,k)\exp \left [ - \, k^2\, \frac{i}2 \int_\eta^{\eta+d\eta}d\eta'  \, \kappa \, \right]  \, ,
\label{eq:Kapprox}
\ee
respectively. Notice that while the wavefunction $\psi$ will oscillate rapidly, the potential $V$ will be a much more smooth and slowly evolving function. 
This means that the integration in (\ref{eq:Vapprox}) can often be replaced by a multiplication with $d\eta$ even for not so small $d\eta$.

According to the Baker-Campbell-Hausdorff formula, the full unitary evolution operator will factorize into the two terms given by (\ref{eq:Vapprox}) and (\ref{eq:Kapprox}) as long as the exponents commute. In particular, as long as one of the exponents is small, the resulting error is small. For the numerical 
evolution of the system we hence use
\bea
\label{eq:Uoperators}
\psi(\eta + d\eta) &=& U_K \, U_P \,  \psi(\eta) \, , \nn \\
\quad U_P  &=& \exp \left [ -i \, d\eta  \,\bar V(\eta,x) \right] \, , \nn \\
\quad U_K  &=& \exp \left [ - \frac{i}2 \, \kappa \, d\eta \,k^2 \right] \, .
\eea
This expression can be a good approximation even in the regime where some of the phases are much larger than unity. This allows us to use relatively large timesteps $d\eta$ in particular in the limit $\hbar \to 0$ where $1/\kappa$ becomes large and $\kappa$ becomes small. In this regime, a simple algorithm to determine a valid timestep $d\eta$ is to ensure that $d\eta  \,\max_x\bar V(\eta,x)< \varphi_{\rm max}\ll 1$ is much smaller than unity. We also checked that our results are independent of the choice of $\varphi_{\rm max}$
(see App.\,\ref{sec:angles}).

The numerical complexity of these equations is very limited. A single timestep requires three Fourier transformations. The run time of a simulation scales as $N_t \, N_x \, \log N_x$, where $N_t$ denotes the number of timesteps and $N_x$ the number of grid points.
For the numerical results shown below we used $N_x=16384$, boxsize $L_{\rm box}=20$\,Mpc, and $\kappa=\hbar/(m H)=0.01 \,{\rm Mpc}^2$ unless otherwise specified.

\section{Numerical results \label{sec:results}} 

The aim of the present analysis is to study the formation of halos and their properties in the \SP{} system. 
The final halo in 1+1 dimensional simulations on a static background can typically be characterized by the inner slope $\rho(r)\propto r^{-\gamma}$
 of the density profile, with $\gamma\sim 0.5$, the halo mass and a cutoff beyond which the density is strongly suppressed~\cite{Binney:2003sn}.
Therefore, if the power-law  scaling is assumed and the exponent would be known, the remaining two characteristics of the
final halo can be predicted since in a static simulation total energy 
\be
E = \int \, dx \, \left( \frac{\hbar^2}{2 a^2 m} \nabla \psi^* \nabla \psi + \frac12 m V \psi^* \psi \right) \, ,
\ee
and total DM mass
\be
M = \int \, dx \, m \, \psi^* \psi \, ,
\ee
are conserved.
It turns out that for almost homogeneous initial conditions, the halo is barely small enough to fit into the simulation volume. Hence, in order to avoid onerous boundary effects, we start with a localized DM density distribution. 

\begin{figure}[t]
\begin{center}
  \includegraphics[width=0.75\textwidth]{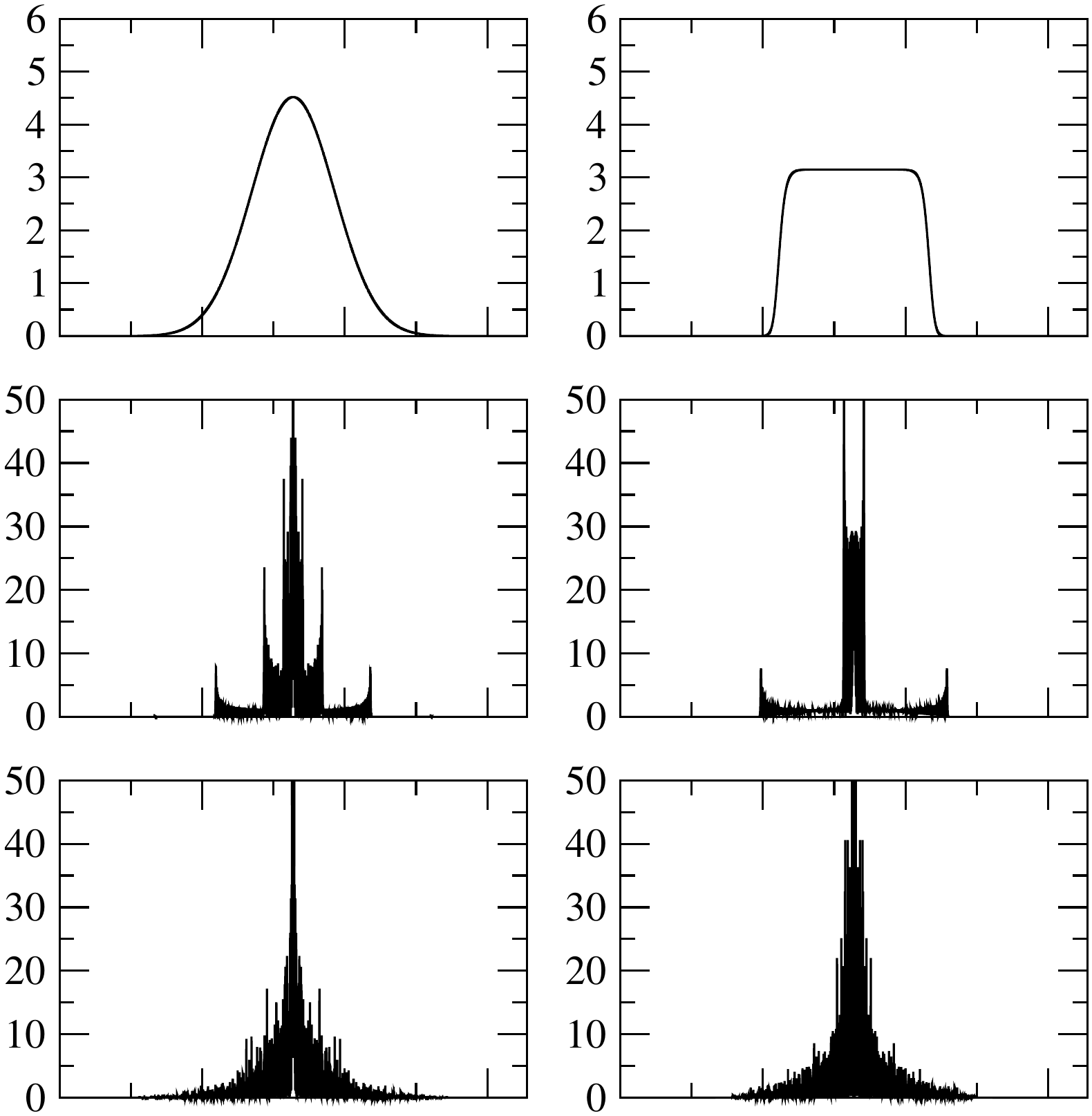}
\end{center}
\caption{\label{fig:static}%
\small The plot shows the DM density at different times in a static simulation without expansion. 
The left plot shows a Gaussian initial distribution while the right plots show the box-like initial distributions. The two upper/middle/lower plots show the simulation at $\eta = 0$, $\eta = 10/3$ and $\eta = 100/3$. The shown region covers the full box.
}
\end{figure}

To be specific, we use two different initial conditions. In the first case, we use a Gaussian distribution %
\be
\rho_{\rm Gauss} \propto \exp\left[- \alpha (x - L/2)^2 \right] \, ,
\ee
that attains the value $e^{-8}$ on the boundary compared with the maximum of the distribution ($\alpha = 32/L^2$). 
This Gaussian curve is narrow enough to fit into the whole simulation after a steady state is reached. As a second initial condition we use a box-like shape
\be
\rho_{\rm box} \propto \tanh \left[ \beta (x + L/6) \right] + \tanh \left[ \beta (L/6 - x) ) \right] \, .
\ee
The box fills initially one third of the simulation volume with an inverse wall thickness of $\beta = 64/L$. The initial densities $|\psi|^2$ for both cases
are normalized to unity and shown in the top left and right panel of Fig.\,\ref{fig:static}, respectively.

In both cases, shortly after the first shell crossing, the system develops strongly fluctuating features. 
For $|\psi|^2$, these are indicated in the lower panels of Fig.\,\ref{fig:static} for two different times and the two initial conditions, respectively. 
Our interpretation in terms of wave packets is the following: Due to our initial conditions, the wave packets are highly correlated and form one coherent field. Once the first shell-crossing occurs, the non-linearities lead to a strong decoherence of the different wave packets that leads to the oscillatory features in the wave function. In order to make the connection to the real phase space distribution of the DM particles, some form of smoothing or coarse-graining is required~\cite{Widrow:1993qq,Davies:1996kp, Uhlemann:2014npa}. This leads to a hierarchy of length scales 
\be
a_{\rm grid} \ll \lambda_{\rm de \, Broglie} \ll \Delta x \ll L_{\rm physics} \ll L_{\rm box} \, .
\ee
Here, $a_{\rm grid}$ is the separation of grid points, $\lambda_{\rm de Broglie}$ denotes the de Broglie wavelength of the wave packets, that is given by the uncertainty relation. Next, $\Delta x$ is the smoothing scale that is applied, while $L_{\rm physics}$ is the length scale of the observables one is interested in. Finally, $L_{\rm box}$ is the box size of our simulation. An added benefit of the \SP{} system is that gravitational softening is already automatically implemented by the wave equations~\cite{Widrow:1993qq}.

To be specific, we use for smoothing the Husimi representation that was already used in~\cite{Widrow:1993qq} 
\be
\psi_H(x, p) = \int dy \, K_H(x,y,p) \, \psi(y) \, ,
\ee
with the Husimi kernel
\be
K_H(x,y,p) = \frac{\exp \left[ -\frac{(x-y)^2}{4 \Delta x^2} - \frac{i}{\hbar} p \, y\right]}
{(2 \pi \hbar)^{1/2} (2\pi \Delta x^2)^{1/4}} \, .
\ee
The coarse-grained particle distribution function is then simply given by 
\be
\label{eq:fH}
f_H(x,p) = |\psi_H(x,p)|^2 \, .
\ee
This expression can then be used to evaluate the smoothed phase space distribution function of the system. 
In the analysis of the higher moments we actually use a Gaussian smoothing on (\ref{eq:defrho}) through (\ref{eq:defsigma}) which is simpler numerically.

\subsection{Higher order cumulants}

\begin{figure}[t]
\begin{center}
  \includegraphics[width=0.7\textwidth]{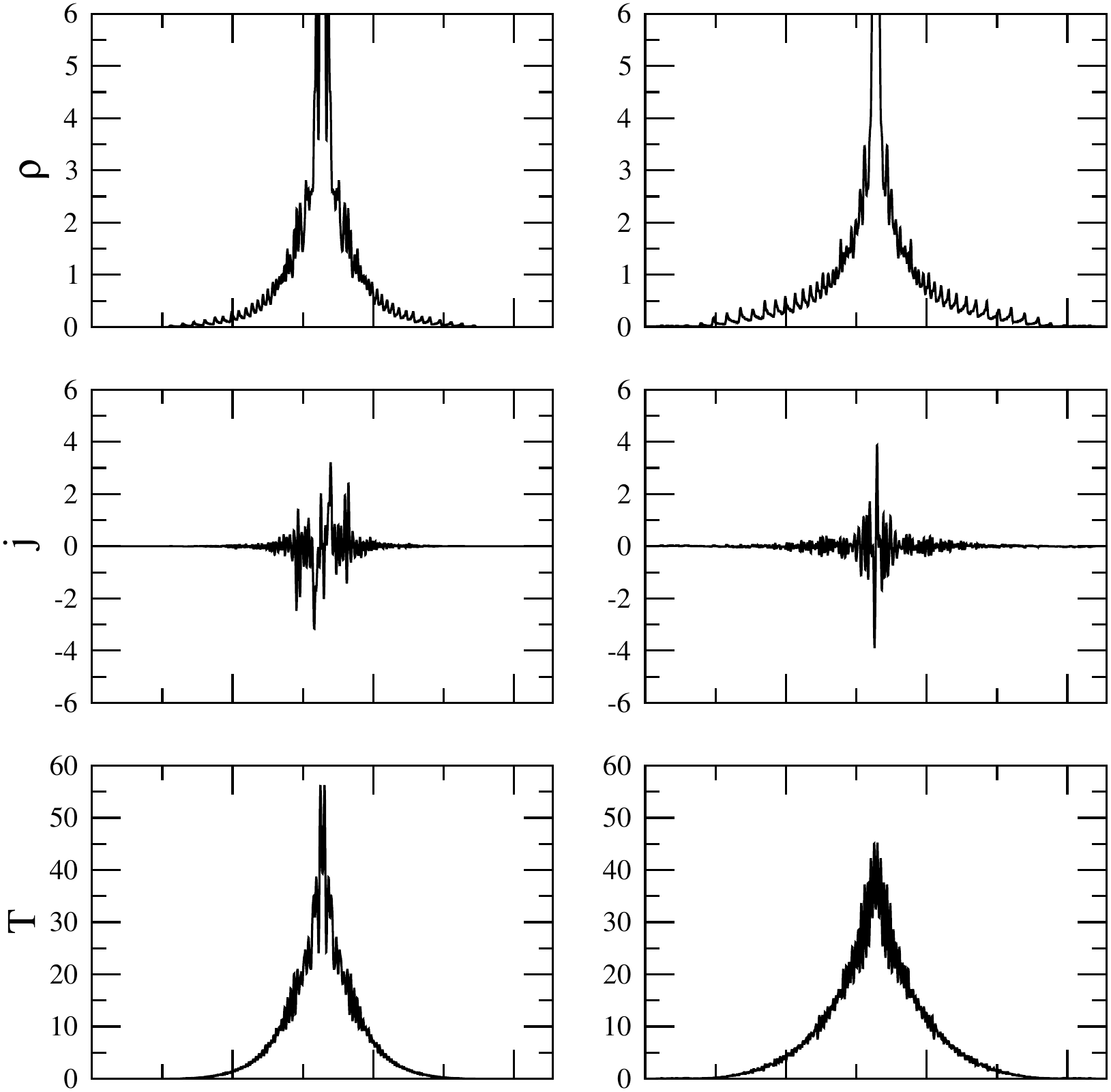}
\end{center}
\caption{\label{fig:moments}%
\small The plot shows the zeroth ($\rho$), first ($j$) and second ($T$) moments of the DM distribution function (from top to bottom) at late time ($\eta=100/3$). 
The left (right) panel corresponds to the Gaussian (box-like) initial distribution. The first moment ($j$) tends to zero at late time as is expected from a steady-state solution. The lines are smoothed over 32 grid points (1/512 of the volume).
}
\end{figure}

At first sight it is surprising that the wave function (that only contains two real degrees of freedom) should be able to represent the full dynamics. After all, the dust model contains two degrees of freedom but does not carry any information about the cumulants of order larger or equal to two. This conundrum is resolved by the fact that the observables are only obtained after smoothing/coarse-graining over a scale that is significantly larger than the typical size of a wave packet. Thus, the fact that each coarse-grained cell contains a significant number of wave packets makes the prediction of higher order cumulants viable - just as for N-body simulations.

Figure~\ref{fig:moments} shows the higher moments of the simulation for the Gaussian and box-like initial conditions for the moments defined in (\ref{eq:defrho}) to (\ref{eq:defsigma}).
Naively, one would expect that the connected pieces of the higher moments scale with higher orders of the parameter $\hbar$~\cite{Uhlemann:2014npa}, because the
\SP{} equation in the Madelung representation approaches the fluid equations for the 
dust-model with vanishing quantum pressure for $\hbar\to 0$. For example, the connected part 
of the second moment, i.e.~the velocity dispersion, is given by
\be\label{eq:sigma}
\sigma = \frac{\hbar^2}{4}\left(\left(\frac{\nabla \rho}{\rho}\right)^2-\frac{\nabla^2 \rho}{\rho}\right)\;.
\ee
However, the argument above has a loop-hole, which can be traced back to the singular behavior of the quantum pressure term for $\rho\to 0$,
that invalidates the naive scaling analysis in terms of $\hbar$.
As already noticed in \cite{Uhlemann:2014npa} the \SP{} system resolves an ambiguity in the dust model that occurs whenever the density $\rho$ approaches zero. In these points, the Madelung representation (\ref{eq:Madelung}) is not faithful since the angle that encodes the velocity field is ambiguous while the wavefunction is still well-defined.

In particular, the scaling  $\propto \hbar^2$ in Eq.\eqref{eq:sigma} is compromised by the fluctuations in the higher derivatives of $\rho$
that grow when $\hbar$ becomes smaller. Therefore, it is possible that, after smoothing, both effects compensate each other and $\sigma$ approaches a finite limit for $\hbar\to 0$,
as one would expect for a steady-state halo.

\begin{figure}[t]
\begin{center}
  \includegraphics[width=0.9\textwidth]{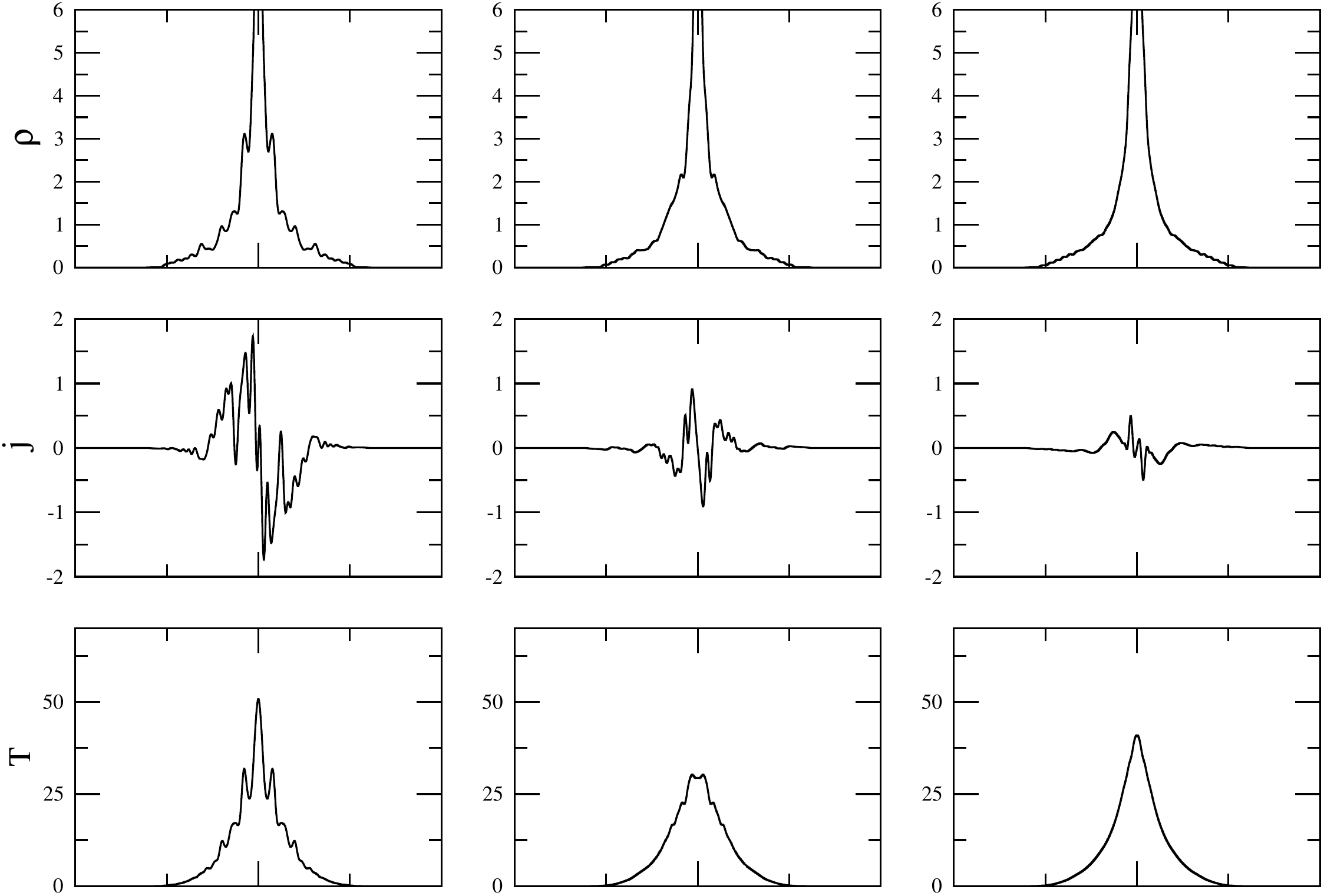}
\end{center}
\caption{\label{fig:hbar}%
\small The plot shows the first three moments $\rho,\, j,\, T$ (top to bottom) for different values of $\kappa=\hbar/(mH)$ 
(left to right: $0.04$, $0.02$, $0.01 \, {\rm Mpc}^2$). For all three values the simulation is already close to virialization. 
The residual of the first moment (second row) is partially due to the finite time of the simulation  ($\eta = 33\frac13$) and partially due to the finite $\hbar$. All plots are smoothed over $1/128$ of the total volume. 
}
\end{figure}

Figure \ref{fig:moments} shows the higher moments for both simulations with a static universe. While the first moment (second row)  $j=u\rho$, that is proportional to the bulk velocity field $u$, vanishes up to effects that are due to finite time and finite $\hbar$, the second moment $T=(\sigma+u^2)\rho$, that involves the velocity dispersion $\sigma$, does not. We checked that when varying $\hbar$ the (smoothed) second moment indeed approaches a finite limit. This indicates that the \SP{} system can account for the generation of higher moments in the distribution function, with a well-defined limiting behavior for $\hbar\to 0$. This limit is illustrated  in Fig.~\ref{fig:hbar}, where we display the first three moments for box-like initial conditions and three different values for $\hbar$. For relatively large values of $\hbar$, the size of the individual wave packets is larger than the smoothing scale $\Delta x$. Accordingly, additional features are visible in the different moments, especially towards the halo center. As mentioned before, the first moment is expected to vanish at late times for $\hbar \to 0$. The reduction for $\hbar\to 0$ can be observed in the second row of Fig.~\ref{fig:hbar}. The residual of the first moment is partially due to the finite time of the simulation and partially due to the finite $\hbar$.

Another way to state the above findings is the following: In the dust model, higher order moments (starting from the second moment) vanish at all times if they are identical to zero in the beginning. In the \SP{} system, on the other hand, higher order moments are generated by the equations of motion even if the density and velocity fields behave close to the dust model initially. This behavior is the one expected from the full Vlasov-Poisson system, and observed also in N-body simulations.
An obvious question is if the second moment of the distribution function is consistent with virialization. This is discussed in the next section in the context of phase space distributions.

\subsection{Phase space distributions}

\begin{figure}[t]
\begin{center}
  \includegraphics[width=0.35\textwidth]{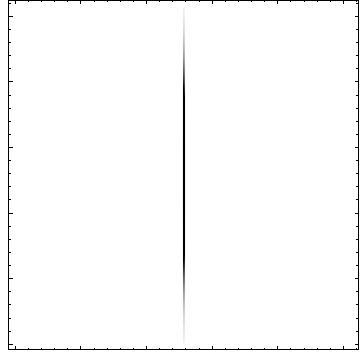}
  \includegraphics[width=0.35\textwidth]{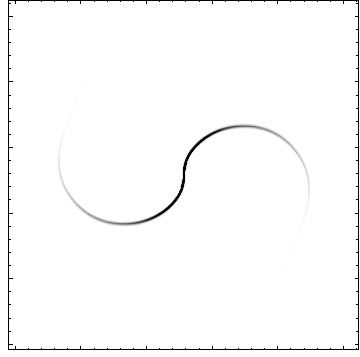}
  \includegraphics[width=0.35\textwidth]{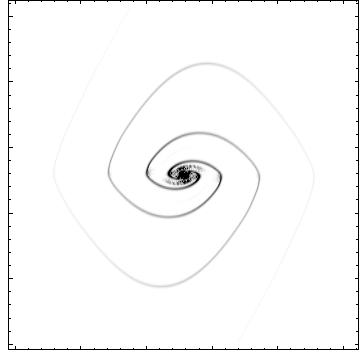}
  \includegraphics[width=0.35\textwidth]{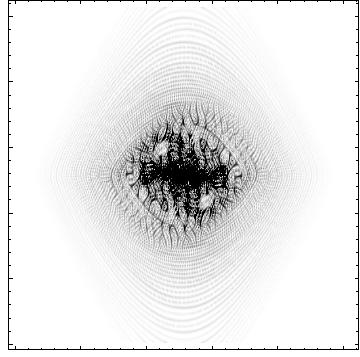}
\end{center}
\caption{\label{fig:pdfs}%
The plot shows the particle distribution functions for the Gaussian initial conditions at the times $\eta = \{ 0,1,3\frac13,33\frac13\}$. The distributions are smoothed over 32 grid points (1/512 of the volume) in coordinate and momentum space. In the first three plots the values are clamped into $[0,1]$ while in the last plot they are clamped into $[0,0.1]$. 
}
\end{figure}

As another benchmark of the \SP{} system we use the smoothed phase space distributions. 
In fact, we do not show the Husimi distribution $f_H(x,p)$ as defined in Eq.~(\ref{eq:fH}) but 
smooth the Wigner transform in $x$- and $k$-space with a Gaussian that is narrower than for the corresponding
Husimi distribution. 
Figure \ref{fig:pdfs} shows four snap shots of the phase space distribution at four different times. 

The pictures resemble the process of phase mixing and violent relaxation \cite{LyndenBell:1966bi}. The strong time-dependence in the gravitational potential makes it possible that matter is transported towards the inner parts of the halo. A quantitative prediction of this scheme is that the time it takes for the overall approach to a stationary halo should be of the same order as the orbital time of the system. This is readily observed in our simulations. The first shell crossing happens at around $\eta \simeq 1$ and the orbital time is in these units around $\simeq 4$. The time for almost complete virialization seems to be a few times longer than this.
The phase space distributions show the folding sheets that are expected from N-body simulations.

\begin{figure}[t]
\begin{center}
  \includegraphics[width=0.65\textwidth]{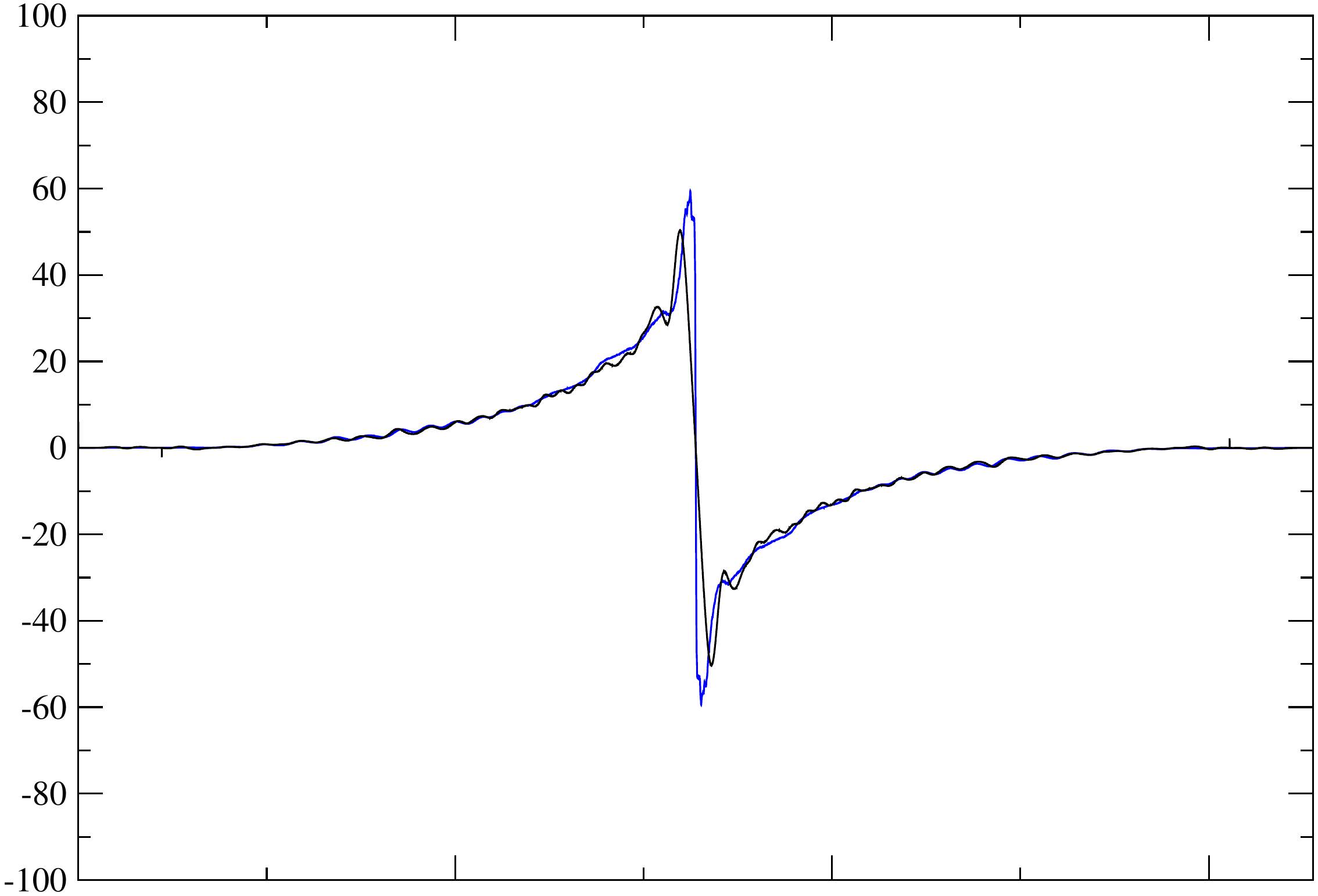}
\end{center}
\caption{\label{fig:virial}%
\small The plot shows the combination $-\nabla \bar V(x) \, \rho(x)$ (in blue) and the first derivative of the second moment $\nabla T(x)$ (in black), both smoothed over $1/128$ of the total volume. These quantities should coincide in case the phase space density does only depend on $\sqrt{V + p^2/2}$. Aside the center of the halo, the agreement is quite good at late times ($\eta = 33\frac13$).
}
\end{figure}

\begin{figure}[pt]
\begin{center}
\includegraphics[width=0.7\textwidth]{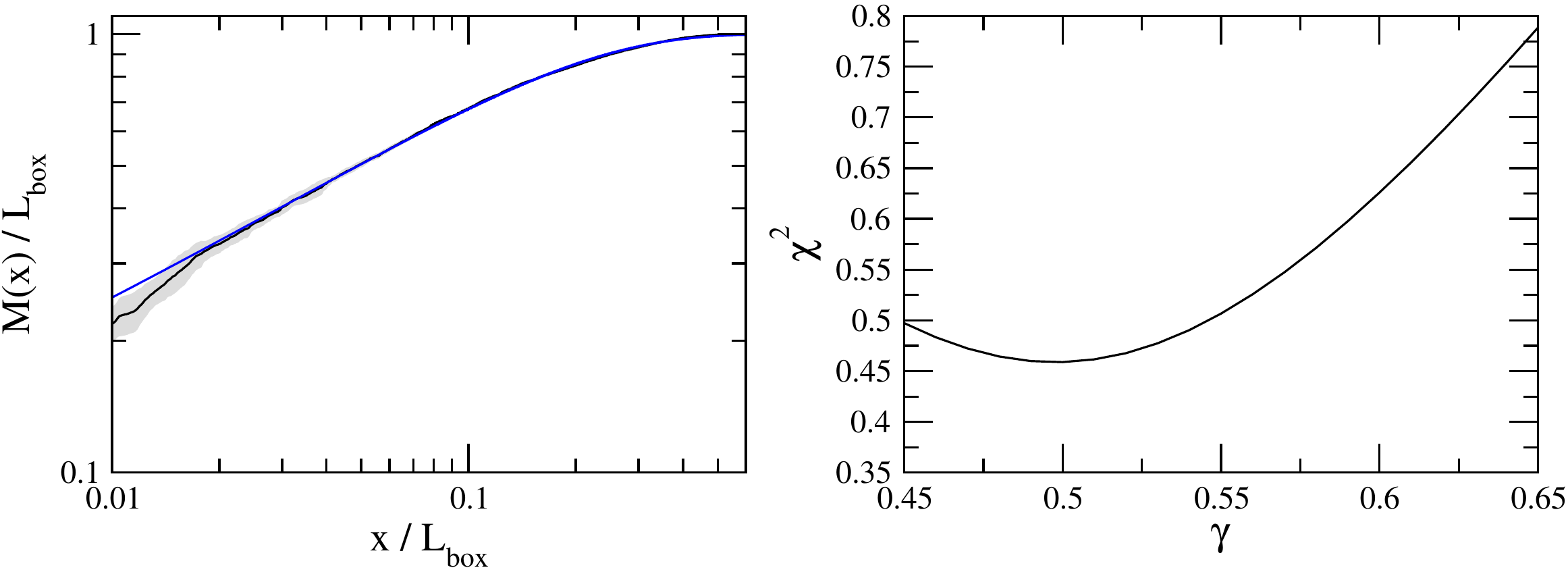}
\includegraphics[width=0.7\textwidth]{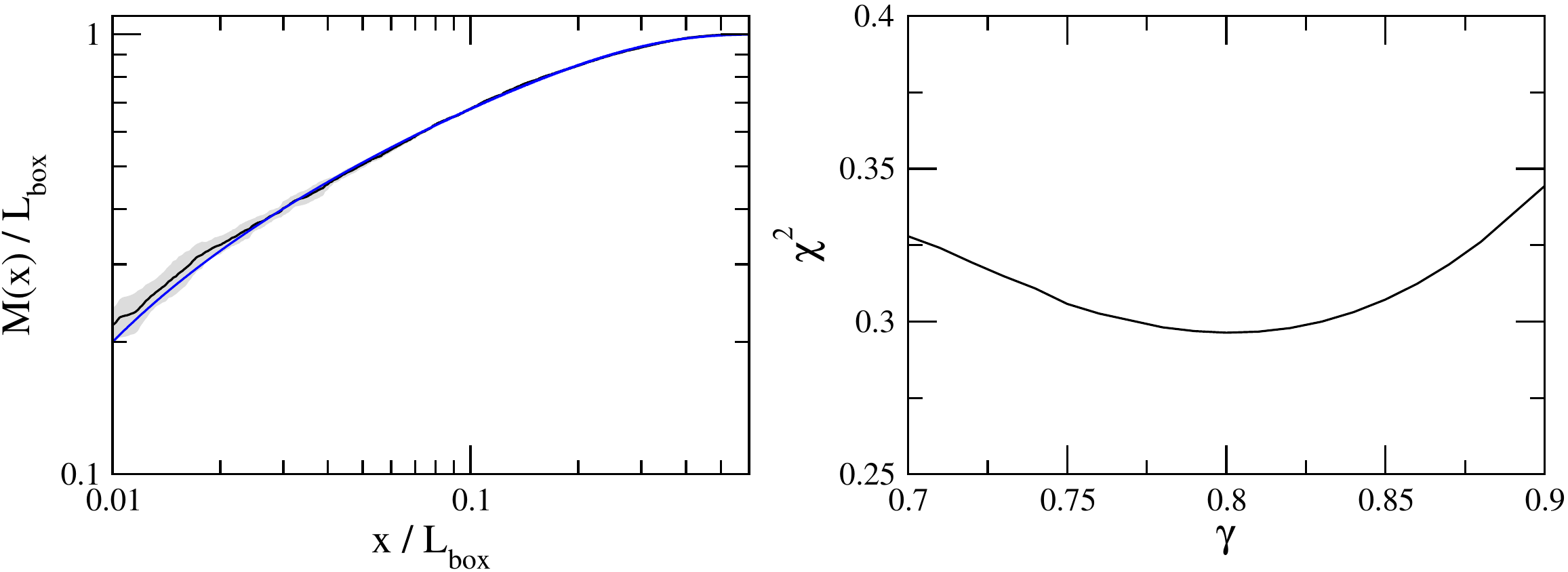}
\includegraphics[width=0.7\textwidth]{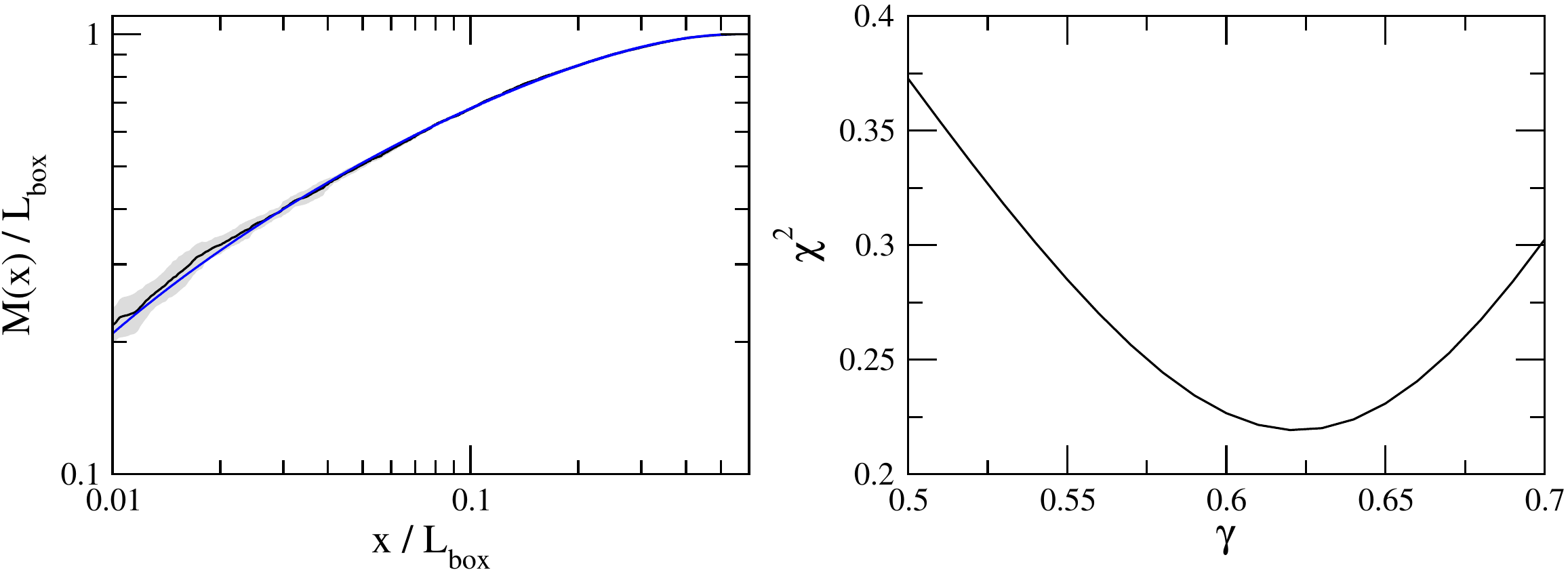}
\end{center}
\caption{\label{fig:scaling}%
\small The three rows show the three fits A, B and C (from top to bottom) for the halo density profile. In the left panels, the integrated density profile $M(x)$ (black line) results from averaging over five different times at $\eta = \{96 .. 100\}$. The shaded region denotes the variance of this average. The blue line denotes the best fit. The right panels show $\chi^2$ as a function of $\gamma$ while minimizing over all other parameters of the fit.
}
\end{figure}

In case the smoothed wave function is stationary, the phase space distribution $f_H(p,x)$ should only be a function of the total energy $(\frac12 p^2 +V(x))^{1/2}$. For the phase space distribution one then finds the relations
\be
\nabla f_H = \nabla V  \, \partial_E \, f_H = \nabla V \frac{1}{p} \partial_p \, f_H \, ,
\ee
and hence for the moments the relation
\be
\nabla \, T(x) = - \nabla \bar V(x) \, \rho(x) \, .
\ee
This relation is displayed in Fig.~\ref{fig:virial} and it holds at late times quite well.
This is actually surprising, since the profile still shows features that indicate that the system is not very close to its equilibrium state. We checked that these features are not due to finite grid size or finite $\hbar$. These features should be  further damped by phase mixing and violent relaxation.

\subsection{Halo density profile}

In this section, we study the density profile of the halo at late times for the different initial conditions.
As we will see, not all initial conditions approach the same asymptotic behavior. 

While the density profile $\rho(x)$ features a strongly fluctuating behavior, the smoothed profile, with fixed smoothing length,
converges to a well-defined halo profile for $\hbar\to 0$. Alternatively, we consider the integrated profile $M(x) = \int_0^x d\tilde x \rho(\tilde x)$,
which we find to converge to a well-behaved functional form for $\hbar\to 0$ even without applying any smoothing.

We employ three different families of functions $\rho(x)$ to fit the corresponding integrated density profile $M(x)$. 
Table \ref{tab:fits} summarizes the fitting functions we use. The first one, fit A, is the one already used for N-body simulations in \cite{Schulz:2012jd}. 
\begin{table}
\begin{center}
\begin{tabular}{| l c |}
\hline
  fit A: & $\rho(x) \propto |x|^{-\gamma} \, \exp( - (|x/x_0|)^{(2-\gamma)})$ \\
\hline
  fit B: & $\rho(x) \propto |x|^{-\gamma} \, \exp( - (|x/x_0|)^\delta)$  \\
\hline
  fit C: & $\rho(x) \propto |x|^{-\gamma} \, \theta(x_0 - x) $  \\
\hline
\end{tabular}
\end{center}
\caption{\label{tab:fits}%
\small The different fits used  for the density profile $\rho(x)$.
}
\end{table}

Shallow initial conditions reached some universal final distribution.
At the same time, the authors of~\cite{Schulz:2012jd} report that this distribution could not be reached with steeper initial conditions. The interpretation of the findings is in essence that the system can transport matter towards the center of the halo using phase mixing and violent relaxation. Transporting matter from the middle of the halo to the edges seems much more inefficient and steep initial conditions persist for long time in the halo profile. This agrees with our findings. In the case of an initial Gaussian distribution, the final halo is considerably shallower than with box-like initial conditions.

The density profile obtained from our numerical solution of the \SP{} equations is compared to the various fitting functions in Fig.\,\ref{fig:scaling} (left panel).
When applying fitting formula $A$, our results are consistent with the findings in~\cite{Schulz:2012jd} ($\gamma\simeq 0.47$) but also with a scaling $\gamma \simeq 0.5$~\cite{Binney:2003sn}. Our simulations do not encompass a wide enough range to decide on this issue. In addition, we note that the best-fit values
for $\gamma$ differ when applying fit B or C, while the quality of the fit is comparable or even slightly better (see right panel of  Fig.\,\ref{fig:scaling}).
For small but finite $\hbar$, our results also indicate a flattening in the innermost region $x \lesssim \sqrt{\hbar/(m H)}=0.1$~Mpc, as expected \cite{Hu:2000ke, Peebles:2000yy} (these scales are not shown in Fig.\,\ref{fig:scaling}). Altogether, we find that the halo density profile obtained via the \SP{} equations is consistent with $N$-body results for cold dark matter~\cite{Binney:2003sn,Schulz:2012jd} in the limit $\hbar\to 0$.

\section{Discussion \label{sec:disc}} 

We presented solutions for the \SP{} system of halo formation in 1+1 dimensions.
Often the \SP{} system is thought to be an extension of the dust model, including quantum pressure,
 via the Madelung representation, that would naively reduce to pressureless perfect fluid equations for $\hbar\to 0$. However, the latter description
becomes singular once shell-crossings occur. The \SP{} equations themselves are free of such singularities, and therefore the limit $\hbar\to 0$ becomes non-trivial
in the multi-streaming regime.

Our findings provide quantitative support for the hypothesis that the \SP{} system is rather an N-body double or even an independent sampling of the full Vlasov-Poisson system. This is very much in the spirit of the original work~\cite{Widrow:1993qq}.
This hypothesis is supported by several findings. First, the \SP{} system allows solutions
coming close to those of N-body simulations and going beyond its interpretation in terms of the dust model by generating higher cumulants even for cold initial conditions. Second, observables such as the first few moments of the distribution function (i.e.~density, velocity and velocity dispersion) converge in the limit $\hbar \to 0$ when averaged over arbitrarily small, but finite regions. Third, the velocity dispersion is related to the gravitational potential in a way that is consistent
with virialization after a few orbital times. Finally, the halo density profile exhibits a power-law behavior with the exponent $\gamma \simeq 0.5$. 
These findings are in disagreement with the dust model but coincide with N-body simulations.

\subsubsection*{Acknowledgements}

We thank Oliver Hahn, Michael Kopp, Rafael Porto and Cora Uhlemann for useful discussions.
We acknowledge support from the Excellence Cluster
``Origin and Structure of the Universe'' at Technische Universit\"at M\"unchen.
We also acknowledge support by the German Science Foundation (DFG) within
the Collaborative Research Center (SFB) 676 Particles, Strings and the Early Universe.


\appendix

\section{Dependence on the time step size \label{sec:angles}} 

\begin{figure}[h]
\begin{center}
  \includegraphics[width=0.9\textwidth]{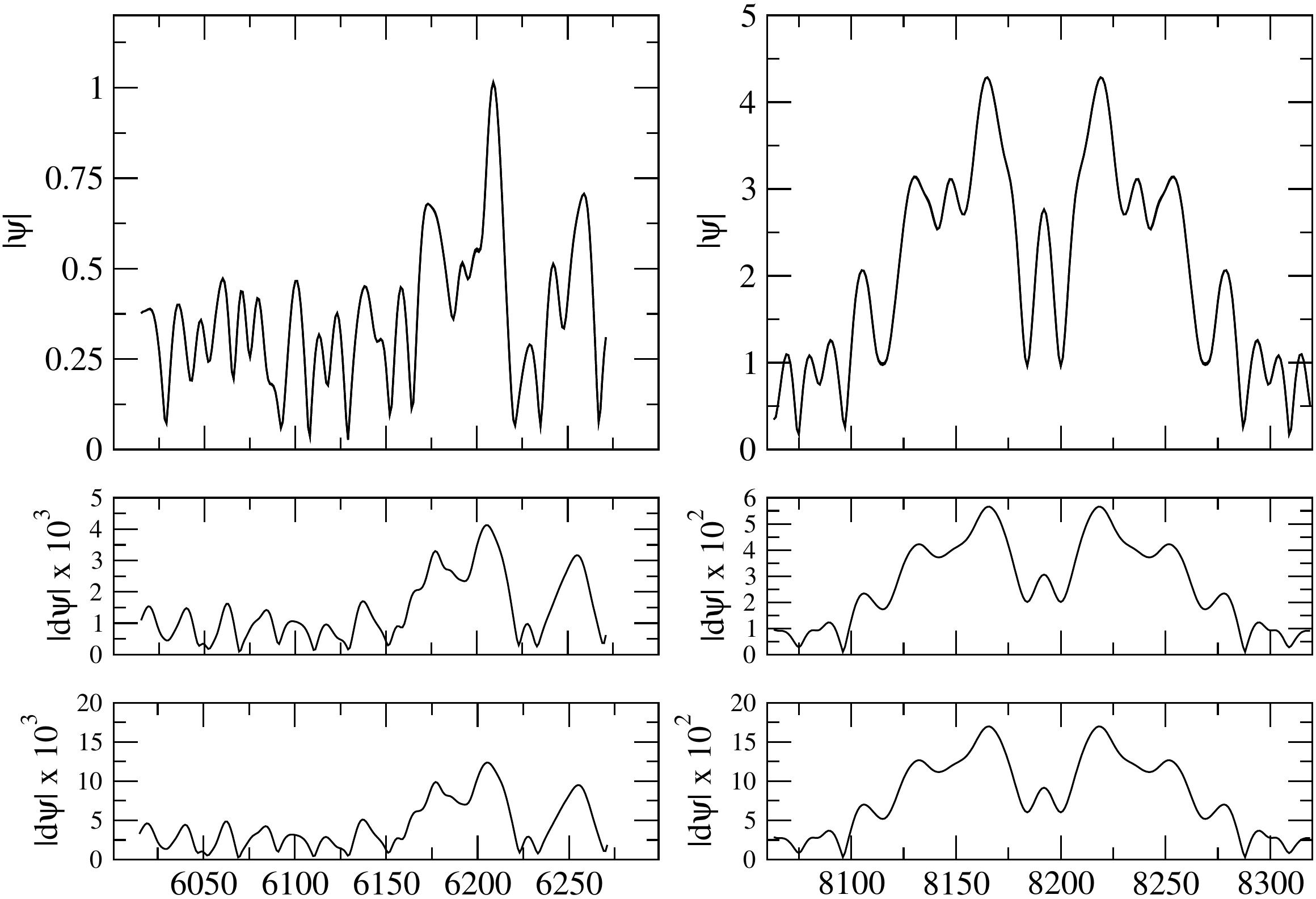}
\end{center}
\caption{\label{fig:angle}%
\small The plot shows the absolute value of the wavefunction $|\psi|$ for three different choices of the maximal phase in the simulation $\{\varphi_{\rm max}^A,\varphi_{\rm max}^B,\varphi_{\rm max}^C\} = \{0.01,0.02,0.04\}$. The three curves basically lie on top of each other. For better visibility, the two lower panels show the difference, $|\psi_A - \psi_B|$ and $|\psi_A - \psi_C|$. The deviations in the middle of the halo (right panel) are somewhat larger than towards the edge (left panel). Both results are obtained for $\eta = 30$ and apparently the error for the set $A$ due to the finite time step is below the percent level.
}
\end{figure}

In this appendix we present some results concerning the convergence of the wavefunction with respect to the maximal phase $\varphi_{\rm max}$ in the unitary operators in (\ref{eq:Uoperators}). When the maximal phase is reduced below $\varphi_{\rm max} \sim 0.01$, changes in the wave function are barely noticeable and are of order $10^{-2}$, see Fig.~\ref{fig:angle}. Accordingly, changes in the coarse-grained observables are even further reduced.

\end{document}